\documentclass[a4paper,10pt]{article}
\usepackage[utf8]{inputenc}
\title{Supersymmetric Duality in Superloop Space }
\author{ Mir Faizal$^1$ and Tsou Sheung Tsun$^2$ \\
$^1$Department of Physics and Astronomy, \\  University of Waterloo,   Waterloo,\\
Ontario N2L 3G1, Canada \\
$^2$Mathematical Institute, University of Oxford,\\ Andrew Wiles
Building, \\Radcliffe Observatory Quarter, Woodstock Road,
\\ Oxford
OX2 6GG, United Kingdom 
 }
 \date{}
\begin{document}
\maketitle
\begin{abstract}
In this paper we constructed  superloop space duality  
 for a four dimensional supersymmetric Yang-Mills theory with $\mathcal{N} =1$ supersymmetry. 
This duality reduces to the ordinary loop space duality for the ordinary Yang-Mills theory. 
  It also reduces to the  
Hodge duality for an abelian gauge theory. 
Furthermore, the electric charges, which are the sources in the original  theory,
appear as monopoles in the dual theory. Whereas,  the magnetic charges, which appear as monopoles 
in the original theory, become  sources in the dual theory.  
\end{abstract}
\section{Introduction}
An important concept in the electromagnetism is the existence of the 
Hodge duality. The symmetry and topological concepts inherent in field theories
 have been analysed using this duality \cite{1a}-\cite{ax11}. 
 In fact, this  duality has been thoroughly studied and many interesting physical consequences  arising from this duality have also been 
analysed \cite{ax22}-\cite{2a}.
 It is known that electrodynamics is dual under Hodge star operation, $^*F_{\mu\nu}= -\epsilon_{\mu\nu\tau\rho}F^{\mu\nu}/2 $. 
This is because the field tensor for pure electrodynamics, $F_{\mu\nu} = \partial_\nu A_\mu - \partial_\mu A_\nu $, satisfies, 
$\partial^\nu F_{\mu\nu} =0 $.  This field tensor also satisfies the Bianchi identity,  ${\partial^\nu}^* F_{\mu\nu} =0 $. This 
field equation for pure electrodynamics can be  interpreted
as the Bianchi identity for $^*F_{\mu\nu}$, because  the Hodge star operation is reflexive, $^*(^*F_{\mu\nu }) = -F_{\mu\nu}$. So,  we can 
express $^* F_{\mu\nu}$,   in terms of  a dual potential, $\tilde A$, such that $^* F_{\mu\nu} = \partial_\nu \tilde A_\mu - \partial_\mu 
\tilde A_\nu $.    It has also been known that 
  the existence of magnetic monopoles is
equivalent to (electric) charge quantization which in turn is
equivalent to the electromagnetic gauge group being compact (i.\ e.\
$U(1)$) \cite{Dirac}. 
However,  a non-abelian version of this duality and
its consequences for non-abelian monopoles  
can only be analysed in the framework 
of loop space  \cite{dual}-\cite{dual1}. 

For Yang-Mills theory, the field tensor,  $F_{\mu\nu} = \partial_\nu A_\mu - \partial_\mu A_\nu  + ig [A_\mu, A_\nu]$, again satisfies, 
$D^\nu F_{\mu\nu} =0 $, where $D_\mu = \partial_\mu - ig A_{\mu}$ is the covariant derivative. It also  
satisfies the Bianchi identity,  ${D^\nu}^* F_{\mu\nu} =0 $. However,  now  this
does not imply the existence  of a  dual potential because the covariant derivative in the Bianchi identity involves the potential 
$A_\mu$ and not some dual potential $\tilde A_\mu$, appropriate to $^* F_{\mu\nu} =0 $. 
In fact, it has been demonstrated that in certain cases no such  solution for such a 
dual potential exist even for the ordinary Yang-Mills theory \cite{dual}-\cite{dual1}. 
Thus, the Yang-Mills theory is not dual under the Hodge star operation. However, it is
possible to construct a generalized duality transformation for the ordinary Yang-Mills 
theory in the  loop space, such that for the abelian case, 
it  reduces to the Hodge star operation \cite{pq1}-\cite{pq2}.

This duality has been used for studding   't Hooft's  order-disorder parameters \cite{1t}. 
For any two spatial loops $C$ and $C'$ with the linking number $n$ between them, 
and the gauge  symmetry generated by the gauge group $su(N)$, the    order-disorder 
parameters satisfy, $A(C) B(C') = B(C') A(C)  exp (2\pi in/N)$. The magnetic flux through $C$ is
measured by $A(C)$. So,  it also creates an electric flux along $C$ and is thus expressed in terms 
of the potential $A_\mu$. However, $B(C)$ measures the electric flux through $C$, and thus creates 
magnetic flux along $C$. So, it can only be expressed in terms of the dual potential  $\tilde A_\mu$
\cite{d}-\cite{1p}. 
A Dualized Standard Model 
has also been constructed using this duality \cite{d}-\cite{d0}. 
In the Standard Model the fermions of the same type but different generations
have widely different masses. The CKM matrix is also not an identity matrix and the off-diagonal
elements of the CKM matrix vary in different magnitude \cite{cmk}. 
These facts can be explained using the Dualized Standard Model \cite{d1}-\cite{d2}.
In fact, even the Neutrino oscillations \cite{no},
and the Lepton transmutations \cite{lt}, have been studied in the Dualized Standard Model. 
 Polyakov loops have been used for deriving this duality in non-abelian gauge theories \cite{p1}.
In mathematical language Polyakov loops are the holonomies of closed
loops in space-time. In fact, in the physics literature they are called Dirac phase factors. 
 Even though they are defined via parameterized
loops in space-time, they are independent of the parameterization
chosen.  They are gauge group-valued functions of the
infinite-dimensional loop space.
The main difference between a Polyakov loop and a Wilson loop is that in the Wilson loop a trace is taken and 
no such trace is taken in the Polyakov loop \cite{p1}.  
Thus, the Polyakov loops are by definition elements of the gauge group. 
It may be noted that Wilsons loops for super-Yang-Mills theory with
$\mathcal{N} =4$ supersymmetry has also been constructed in superspace formalism \cite{ps}.
The Polyakov loops for three  and four dimensional supersymmetric Yang-Mills theories with 
$\mathcal{N} =1$ supersymmetry have also been studied \cite{pqaa}-\cite{pq1a}.  
   In this paper we will derive a supersymmetric duality for 
 the four dimensional supersymmetric Yang-Mills theory in the  Wess-Zumino gauge.

\section{Superloop Space}
In four dimensional gauge theories with $\mathcal{N} =1$
supersymmetry,  we can construct a covariant derivative 
$ \nabla_A = D_A - i \Gamma_A$, where $D_A = ( \partial_{a\dot{a}}, D_a, D_{\dot{a}})$
and $\Gamma_A = (\Gamma_{a\dot{a}},\Gamma_a, \Gamma_{\dot{a}} )$ \cite{1001}. 
Furthermore, the Bianchi identity can now be written as 
$ [\nabla_{[A}, H_{BC)}\} =0, 
$ where
$
H_{AB}=  [ \nabla_A, \nabla_B \} = T^C_{AB}\nabla_C - i F_{AB} 
$. 
Thus, again for a supersymmetric Yang-Mills theory, no dual
potential can be constructed.
However, as it is possible to derive a duality for the
ordinary Yang-Mills theory in loop space, we will derive a 
duality a supersymmetric Yang-Mills theory in superloop space formalism.
We will derive our results in Wess-Zumino gauge, and impose the constraint 
$F_{a\dot{a}} = F_{ab} = F_{\dot{a}\dot{b}} =0$.
Now can  we  
define $ \xi (s)= (\sigma^\mu \xi_\mu(s))^{a\dot{a}}\theta_{a}\theta_{\dot{a}} + 
\xi^a (s)\theta_a + \xi^{\dot{a}} (s)\theta_{\dot{a}} $, and so we have 
$\xi^A = (\xi^{a\dot{a}}, \xi^a, \xi^{\dot{a}})$ \cite{1001}.
We have to used this parameterization of the superloop space as we are analysing a 
theory with $\mathcal{N} =1$ supersymmetry in four dimensions  \cite{pq1a}.
This is because for a four dimensional gauge theory with $\mathcal{N} =1$ supersymmetry, the gauge 
fields can be obtained from a super-connection which is given by 
$\Gamma_A = (\Gamma_{a\dot{a}},\Gamma_a, \Gamma_{\dot{a}} )$. The superloop space is constructed 
to analyse this theory, and so we have chosen this particular form of parameterized. 
It may be noted 
that if had considered supersymmetric theories with higher amount of supersymmetry, 
we would have to take additional Grassmann coordinates into consideration. 
In fact, Wilsons loops for such theories have been constructed  by taking these extra Grassmann coordinates 
into consideration \cite{ps}.
This would also occur if we were constructing superloop space in higher dimensions.
Furthermore,  it has been demonstrated that three dimensional superloop space 
requires a small number of  Grassmann coordinates \cite{pqaa}. 
The superloop can now be parameterized by  $\xi^A 
= (\xi^{a\dot{a}}, \xi^a, \xi^{\dot{a}})$, along a curve $C$, 
\begin{equation}
 C : \{ \xi^A (s): s = 0 \to 2\pi, \, \, \xi^A (0) = \xi^A(2\pi)\},  
\end{equation}
where   $\xi^A (0) = \xi^A(2\pi)$ is a fixed point on this curve.
The space of all such super-functions parameterizes the superloop space. 
A   functional on this superloop space can be constructed as  \cite{pq1a}
\begin{eqnarray}
 \Phi [\xi] &=& 
 P_s \exp i \int^{2\pi}_0 \left[ \Gamma^{a\dot{a}} (\xi(s))  \frac{d \xi_{a\dot{a}}(s)}{ds} +  \Gamma^a (\xi(s))  
\frac{d \xi_a(s)}{ds}  +   \Gamma^{\dot{a}} (\xi(s))  
\frac{d \xi_{\dot{a}}(s)}{ds}\right] 
 \nonumber \\  &=&
 P_s \exp i \int^{2\pi}_0  \Gamma^A (\xi(s)) \frac{d \xi_A(s)}{ds}. 
\end{eqnarray}
here  $P_s$ denotes ordering in $s$ 
increasing from right to left and the  derivative in $s$ is  taken from below.
This loop space variable is a scalar superfield from the supersymmetric point of view, and can be projected to component superloops. 
In particular, we have  
 $
[\Phi[\xi]]_| = \phi[\xi]
$, which in Wess-Zumino gauge is given by 
\begin{equation}
  \phi [\xi] =
 P_s \exp i \int^{2\pi}_0 A^{\mu}(\xi(s))\frac{d \xi_{\mu}(s)}{ds}. 
\end{equation}

We can also define the parallel transport from a point $\xi(s_1)$ to $\xi(s_2)$ along path parametrized by $\xi$
as 
 \begin{equation}
 \Phi [\xi: s_1, s_2 ] 
= P_s \exp i \int^{s_2}_{s_1}  \Gamma^A (\xi(s)) \frac{d \xi_A}{ds}
\end{equation}
Now using $\Phi[\xi]$, we can define a gauge Lie algebra valued  $F_A[\xi|s]$ as  
\begin{eqnarray}
 F_A [\xi| s] &=&
 i \Phi^{-1}[\xi]\delta_A (s) \Phi[\xi]
  \nonumber \\
 &=& \Phi^{-1}[\xi: s,0] H^{AB} (\xi (s) )\Phi  [\xi: s,0]\frac{d \xi_B (s) }{d s},  
\end{eqnarray}
where $\delta_A (s) = \delta /\delta \xi^A (s)= 
(\delta /\delta \xi^{a\dot{a}}(s),\delta /\delta \xi^a (s), \delta /\delta \xi^{\dot{a}} (s)))  $.
Here we first followed a path to $s$ and  then turn backwards 
along the same path. Thus, the phase factor for the segment of the superloop beyond $s$ did not contribute and
 $H^{AB}(\xi (s) )$ was obtained because of the  infinitesimal circuit 
generated at $s$.

It is convenient at this stage to define a functional curl and a functional divergence for 
these superloop space variables as 
\begin{eqnarray}
 (\rm{curl}\,  F [\xi|s])_{AB} &=& \delta _A (s) F_B[\xi|s] -
\delta _B (s) F_A[\xi|s], \nonumber \\
\rm{div}\,  F [\xi|s] &=& \delta ^A (s) F_A[\xi|s].
\end{eqnarray}
These superloop variables are highly redundant and have to be constrained by an infinite set of conditions 
which can be expressed by the vanishing of the superloop space curvature  \cite{pq1a},
$G_{AB}[\xi, s] =  (\rm{curl}\,  F [\xi|s])_{AB}
+i [F_A [\xi|s], F_B [\xi|s]]
= 0 $. 
Now we construct $E_A [\xi|s]$ from $F_A [\xi|s]$ as follows, 
\begin{equation}
 E_A [\xi|s] =  \Phi[ \xi: s, 0] F_A[\xi|s]  \Phi^{-1}[ \xi: s, 0], 
\end{equation}
So,  $E_A [\xi|s]$ is obtained from a parallel transport of $F_A [\xi|s]$. 
Thus, $E_A [\xi|s]$ depends only on a segment of the loop  $\xi(s)$ around $s$ and is therefore a segmental variable rather than a full loop variable. However, 
when this segment shrinks to a point, we have $E^A [\xi|s] \to H^{AB} (\xi (s)) d \xi_B (s) /d s$. 
This limit has to be taken only after all the loop operations such as loop differentiation has been performed. This is because all these loop operations require a segment of the loop on which they can operate. 
Now we can define a  functional curl and a functional divergence for 
 $E_A [\xi|s]$ as 
\begin{eqnarray}
 (\rm{curl}\, E [\xi|s])_{AB} &=& \delta _A (s) E_B[\xi|s] -
\delta _B (s) E_A[\xi|s], \nonumber \\
\rm{div}\, E [\xi|s] &=& \delta ^A (s) E_A[\xi|s].
\end{eqnarray}
We first note that 
\begin{eqnarray}
 \delta_A (s') E_B[\xi|s] &=&  \Phi[ \xi: s, 0] [ \delta_A (s') F_B[\xi|s] 
 \nonumber \\ && + i \Theta (s-s') [F_A [\xi|s], F_B [\xi|s]]]\Phi^{-1}[ \xi: s, 0], 
\end{eqnarray}
where $i \Theta (s-s')$ is the Heavisde function. 
So,  the superloop space curvature can now be written as $ G_{AB }[\xi, s] = 
\Phi[ \xi: s, 0] ({\rm{curl}} E [\xi|s] )_{AB} \Phi^{-1}[ \xi: s, 0]$ and thus the 
constraints can be fixed as $({\rm{curl}}E [\xi|s] )_{AB} =0$.

\section{Duality}
For ordinary gauge theories, 
it is possible to construct a duality using loop space formalism, such that it reduces to the 
Hodge star operation for the abelian case \cite{pq1}-\cite{pq2}. In this section we will further 
generalize this duality from a ordinary Yang-Mills theory to a supersymmetric Yang-Mills theory.  
In order to achieve this we define a new  variable 
$\tilde E_A [ \eta|t]$ which is dual to 
$E_A [\xi|s]$. Here $\eta$ is another parameter loop which is parameterized by 
$t$, and    $ \eta (t)= \eta^a (t)\theta_a + (\gamma^\mu \eta_\mu(t))^{ab}\theta_{a}\theta_b  $. 
We have used a different  labels for the parameters of the superloop space, i.e., 
$t$ and $\eta(t)$ instead of $s$ and $\xi(s)$ 
to distinguish the parameters that parameterizing the dual superloop space from 
the parameters that parameterizing the original superloop space. 
So, the parameters $t$ and $\eta(t)$ parameterizing the dual superloop space and the parameters
  $s$ and $\xi(s)$ parameterizing the original superloop space. 
This dual variable is constructed as follows, 
\begin{eqnarray}
 \omega^{-1} [\eta(t) ] \tilde E^A [ \eta|t] \omega [\eta(t)]&=&  
 - \frac{2}{N} \epsilon^{ABCD}\frac{d\eta_B(t)}{dt} \int  D\xi ds E_C [\xi|s] \frac{d\xi_D (s)}{ds}
 \nonumber \\ && \times 
 \left[\frac{d\xi^F (s)}{ds}\frac{d\xi_F(s) }{ds} \right]^{-2} \delta(\xi (s) - \eta(t)), \label{1d}
\end{eqnarray}
where $N$ is a normalization constant. In the tensor $\epsilon^{ABCD}$, the variables $A, B, C, D$ take 
all the possible spinor values i.e., $A = (a\dot{a}, a, \dot{a}), 
B = (b\dot{b}, b, \dot{b}), C = (c\dot{c}, c, \dot{c}), D = (d\dot{d}, d, \dot{d})$.  
Here $ \omega [\eta(t) ] $ is a local rotational matrix which accounts 
for transforming the quantities from a direct frame to the dual frame. 
In the integral $E_C [\xi|s]$ depends on a little segment from $s_-$ to $s_+$, such that the limit $\epsilon \to 0$ is taken only after integration, where $\epsilon = s_+ - s_- $.  As we may need to calculate the loop derivative of $\tilde E^A [ \eta|t]$,
so we regard $\tilde E^A [ \eta|t]$ as a segmental quantity depending on a segment from $t_-$
to $t_+$ and  only  after differentiation the limit $\epsilon' \to 0$ is taken, where 
$\epsilon' = t_+ -t_-$.  This limit is taken
before the limit $\epsilon \to 0$ for the integral. Thus, we can take $\epsilon' < \epsilon$, 
and the $\delta$-function now ensures that $\xi (s)$ coincides from  $s = t_-$ to $s = t_+$ with $ \eta (t)$. After the limit is taken and the segment shrinks to a point, we have   $E^A [\eta|t] \to \tilde H^{AB} (\eta(t)) d \eta_B (t) /d t$. Here $\tilde H^{AB}$ can be constructed from a dual 
potential.  Thus, this superloop space duality implies the existence of 
a dual potential $\tilde \Gamma_A = (\tilde \Gamma_{a\dot{a}},\tilde\Gamma_a, \tilde 
\Gamma_{\dot{a}})$, such that $ [\tilde \nabla_A, \tilde\nabla_B \} =  \tilde H_{AB} $
 where $ \tilde\nabla_A = D_A - i \tilde \Gamma_A$. 

It may be noted that if we used $[\Phi[\xi]]_| = \phi[\xi]$ as the loop space variable, then this duality would reduce 
to the ordinary duality for the  ordinary  Yang-Mills fields. 
So, if we use $[\Phi[\xi]]_| = \phi[\xi]$ as the loop space variable, then we can 
construct $E_\mu [\xi|s]$ from $F_\mu [\xi|s]$, where 
$F_\mu [\xi|s]$ is the loop space connection  corresponding to loop variable $[\Phi[\xi]]_| = \phi[\xi]$, in the Wess-Zumino gauge. 
Then $\tilde E^\mu [ \eta|t]$, which is dual to  $E_\mu [\xi|s]$, is given by
\begin{eqnarray}
 \omega^{-1} [\eta(t) ] \tilde E^\mu [ \eta|t] \omega [\eta(t)]&=&  
 - \frac{2}{N} \epsilon^{\mu\nu \tau \rho}\frac{d\eta_\nu(t)}{dt} \int  D\xi ds E_\tau [\xi|s] \frac{d\xi_\rho (s)}{ds} \nonumber \\ && \times 
 \left[\frac{d\xi^\sigma (s)}{ds}\frac{d\xi_\sigma(s) }{ds} \right]^{-2} \delta(\xi (s) - \eta(t)), 
\end{eqnarray}
If we  let the segmental width of $\tilde E^\mu [ \eta|t]$ go to zero,  then we can write 
\begin{eqnarray}
 \omega^{-1} [x ] \tilde F^{\mu\nu} [x] \omega [x]&=&  
 - \frac{2}{N} \epsilon^{\mu\nu \tau \rho} \int  D\xi ds E_\tau [\xi|s] \frac{d\xi_\rho (s)}{ds} \nonumber \\ && \times 
 \left[\frac{d\xi^\sigma (s)}{ds}\frac{d\xi_\sigma(s) }{ds} \right]^{-2} \delta(x - \xi (s)), 
\end{eqnarray}
Here we first do the integration before taking the limit to zero. Thus,
 in the abelian case, when we take the the limit $\epsilon \to 0$, we obtain \cite{pq2}
 \begin{eqnarray}
 \tilde F^{\mu\nu} [x] &=&  
 - \frac{2}{N} \epsilon^{\mu\nu \tau \rho} \int  D\xi ds F_{\tau \lambda} [\xi(s)]   \frac{d\xi^\lambda (s)}{ds}\frac{d\xi_\rho (s)}{ds}  \nonumber \\ && \times 
 \left[\frac{d\xi^\sigma (s)}{ds}\frac{d\xi_\sigma(s) }{ds} \right]^{-2} \delta(x - \xi (s))
 \nonumber \\ &=& 
 -\frac{1}{2} \epsilon^{\mu\nu\tau \rho} F_{\tau \rho} [x]. 
\end{eqnarray}
Now identifying $\tilde F_{\mu\nu}$ with $^* F_{\mu\nu}$, we obtain the Hodge star operation for ordinary electrodynamics. 
Thus, for the ordinary 
abelian  gauge theory, this duality reduces to the usual Hodge duality. 
\section{Sources and Monopoles}
We will shown in this section that this duality in the superloop space transforms 
the electric charges, which are the sources in the original  theory,
into monopoles in the dual theory. It also transforms  the magnetic charges, which are monopoles  
in the original theory, into  sources in the dual theory.
In order to prove this result, it is useful to first show that this duality is invertible. 
This can be demonstrated by first defining $E^A [ \zeta|u]$ as, 
\begin{eqnarray}
 \omega^{-1} [\zeta(u) ]E^A [ \zeta|u] \omega [\zeta(u)]&=&  
 - \frac{2}{N} \epsilon^{ABCD}\frac{d\zeta_B(u)}{du} \int  D\eta dt \tilde E_C [\eta|t] 
 \frac{d\eta_D (t)}{dt} \nonumber \\ && \times 
 \left[\frac{d\eta^F (t) }{dt}\frac{d\eta_F (t) }{dt} \right]^{-2} \delta(\eta (t) - \zeta(u)), 
\end{eqnarray}
where $\zeta_B (u)$ is a new loop parameterized by $u$.
Now we define $A^A [\zeta (u) ]$ as 
\begin{eqnarray} A^A [\zeta (u) ]&=&
 \frac{2}{N} \epsilon^{ABCD}\frac{d\zeta_B(u)}{du} \int  D\eta dt \omega^{-1} [\eta(t) ]\tilde E_C [\eta|t] 
 \omega [\eta(t) ] \nonumber \\ && \times 
 \frac{d\eta_D (t)}{dt} 
 \left[\frac{d\eta^F (t) }{dt}\frac{d\eta_F (t) }{dt} \right]^{-2} \delta(\eta (t) - \zeta(u))
 \nonumber \\ & =& 
 -\frac{4}{N} 
  \epsilon^{ABCD}\frac{d\zeta_B(u)}{du} \int  D\eta D \xi  dt ds  
 \frac{d\eta_D(t)}{dt} \frac{d\eta^Q(t)}{dt} \nonumber \\ &&\times 
 \left[\frac{d\eta^X (t) }{dt}\frac{d\eta_X (t) }{dt} \right]^{-2}
 \delta(\eta(t) - \zeta (u))E^W [\xi|s]  \nonumber \\ && 
 \times  \frac{d\xi^E(s)}{ds} 
 \left[\frac{d\xi^Y (s) }{ds}\frac{d\xi_Y (s) }{ds} \right]^{-2} \delta(\xi (s) - \eta(t))
 \epsilon_{CQWE}.
\end{eqnarray}
Thus, we obtain,  
\begin{eqnarray}
 \omega^{-1} [\zeta(u) ] E^A [ \zeta|u] \omega [\zeta(u)]&=&  
 - \frac{2}{N} \epsilon^{ABCD}\frac{d\zeta_B(u)}{du} \int  D\eta dt \tilde E_C [\eta|t] 
 \frac{d\eta_D (t)}{dt} \nonumber \\ && \times 
 \left[\frac{d\eta^F (t)}{dt}\frac{d\eta_F(t) }{dt} \right]^{-2} \delta(\eta (t) - \zeta(u)). \label{2d}
\end{eqnarray}
Now identifying $\zeta (u) $ with $\xi(s)$, we obtain the desired result that  
 this duality is invertible. Now if we compare Eq. (\ref{1d}) to Eq. (\ref{2d}), we observe that 
 that Eq. (\ref{1d}) transforms the original superloop space variable to the dual superloop variable, and 
 Eq. (\ref{2d}) inverts that transformation, transforming the dual superloop variable to the original 
 superloop space variable. Hence, this transformation is invertible. 
 
The color electric charge is the source term in the supersymmetric Yang-Mills theory. 
Thus, it can be defined as the  non-vanishing of $\nabla^C H_{BC}$. Alternately, it can also be 
defined as the non-vanishing of $ {\rm{div}}F [\xi |s]$. Furthermore, as
${\rm{div}}E [\xi |s] = \Phi[ \xi: s_1, 0] {\rm{div}}F [\xi |s]] \Phi^{-1}[ \xi: s_1, 0]$, 
so the color electric charge can also be defined as the non-vanishing of 
${\rm{div}}E [\xi |s]$. 
Similarly, as the 
the color magnetic charge is  a monopole in the supersymmetric Yang-Mills theory, it is characterized by non-vanishing of  $G_{AB}[\xi, s]$. So,  
the color  magnetic charge can be defined as the non-vanishing of  $({\rm{curl}} E[\xi|s])_{AB} $. A monopole in the dual theory is also characterized by non-vanishing of $({\rm{curl}} \tilde E[\eta|t])_{AB} $, and  a source in the dual theory 
is defined as the non-vanishing of ${\rm{div}}\tilde E [\eta |t]$. So, under the duality transformation a electric charge in the original theory should appear as a magnetic monopole in the dual theory. So, 
the non-vanishing of ${\rm{div}}E[\xi|s]$ should imply the non-vanishing of 
$({\rm{curl}} \tilde E [\eta|t])_{AB}$. Furthermore, a magnetic monopole in the original theory 
should appear as the source term in the dual theory. So, the non-vanishing of $({\rm{curl}} E[\xi|s])_{AB} $ should imply the non-vanishing of ${\rm{div}}\tilde E [\eta |t]$. 
 Now as $\eta (t)$  coincides with 
$\xi (s)$ from  $s = t_-$ to $s = t_+$, so we can write
\begin{eqnarray}
&&\frac{\delta}{\delta \eta_M (t)} \left(\omega^{-1} [\eta(t) ] \tilde E^A [ \eta|t] \omega [\eta(t)]\right) \epsilon_{MANP}  
 \nonumber \\&=&  
 - \frac{2}{N} \epsilon^{ABCD}\frac{d\eta_B}{dt} \int  D\xi ds \frac{\delta E_C [\xi|s] }{\delta \xi_M (s)}\frac{d\xi_D}{ds} 
 \nonumber \\ && \times 
 \left[\frac{d\xi^F}{ds}\frac{d\xi_F}{ds} \right]^{-2} \delta(\xi (s) - \eta(t)) \epsilon_{MANP}.
\end{eqnarray}
Here we have performed the integration by parts with respect to $D\xi$. 
This expression can be simplified to the following expression, 
\begin{eqnarray}
 \left(\omega^{-1} [\eta(t) ] ({\rm{curl}}\tilde E [ \eta|t]_{AB} \omega [\eta(t)]\right)
 &=& - \frac{1}{N}  \int  D\xi ds 
 \mathcal{A}_{AB} (t, s) {\rm{div}}E[\xi |s]   \nonumber \\ && \times 
 \left[\frac{d\xi^F}{ds}\frac{d\xi_F}{ds} \right]^{-2} \delta(\xi (s) - \eta(t)),
\end{eqnarray}
where 
\begin{equation}
\mathcal{A}_{AB}(t, s) =  \left[\frac{d\eta^C (t)}{d t} \frac{d  \xi^D(s)}{ds}
-\frac{d\eta^D (t) }{d t}\frac{d \xi^C (s)}{ds}\right]\epsilon_{ABCD}.
\end{equation}
Now if ${\rm{div}}E[\xi|s] =0$, then $({\rm{curl}} \tilde E[\eta|t])_{AB}= 0$. 
As the duality is invertible,  we can also show that if 
${\rm{div}}\tilde E[\xi|s] =0$, then $({\rm{curl}}  E[\eta|t])_{AB}= 0$. 
So, an electric charge which is a source in the original theory appears as a monopole in the dual theory, 
and magnetic charge which is a source in the dual theory appears as a monopole in the original theory. 
 
\section{Conclusion}
In this paper we  have analysed a four dimensional pure Yang-Mills theory  with 
$\mathcal{N} =1$ supersymmetry in superloop space formalism. 
We have constructed a generalized duality in superloop space, for this theory. 
Under this generalized duality transformation   the electric charges which appear as sources in the 
original theory become monopoles in the dual theory. Furthermore,  the magnetic charges which appear
monopoles in the original 
theory become  sources
in the dual theory. This duality reduces to the ordinary loop space duality for ordinary Yang-Mills theory.
As the loop space duality
for ordinary Yang-Mills theory reduces to the Hodge star operation in the abelian case, so,  
this generalized duality transformation  
also reduces to the  Hodge star operation for ordinary electrodynamics.

It may be noted that the existence of a duality for ordinary Yang-Mills theory has  many 
interesting physical consequences \cite{1p}-\cite{d2}. It will be interesting to construct a supersymmetric version 
of these results using the results of this paper. Thus, the results of this paper can be used to construct
a supersymmetric Dualized Standard Model.
It will also be interesting to analyse the phenomenological consequences of this model.
The supersymmetric   Standard Model contains a supersymmetric  matter action coupled to 
the supersymmetric gauge theory. Thus, we will need to couple a supersymmetric matter 
action to the supersymmetric gauge  
theory, and then use this formalism to construct a supersymmetric Dualized Standard Model.
In particular,
we expect to have a dual symmetry 
corresponding
to the super-gauge symmetries of the supersymmetric Standard Model.  It will 
also be interesting to generalize the results of this paper to theories with greater amount
of supersymmetry.  The results obtained in this paper can also be used for analysing monopoles in the
ABJM theory \cite{abjm}. 
It may be noted that the supersymmetry of the ABJM theory is expected to get enhanced because of monopole 
operators \cite{enha}-\cite{enha1}. 
Thus, the formalism developed in this paper could find application in the supersymmetry enhancement of
the ABJM theory.

\end{document}